\documentclass[twocolumn,showpacs,preprintnumbers,amsmath,amssymb]{revtex4-1}

\usepackage{graphicx} \usepackage{bm} \usepackage{dcolumn}

\begin{document}
\title{Two-photon echo method for observing electron Zitterbewegung in carbon nanotubes}
\date{\today}
\author{Tomasz M. Rusin$^1$ \ and Wlodek Zawadzki$^2$}
\address{$^1$ Orange Customer Service sp. z o. o., Al. Jerozolimskie, 02-326 Warsaw, Poland\\
         $^2$ Institute of Physics, Polish Academy of Sciences, 02-668 Warsaw, Poland}

\email{Tomasz.Rusin@orange.com}


\begin{abstract}

The phenomenon of Zitterbewegung (ZB, trembling motion) of electrons is
described in zigzag carbon nanotubes (CNT) excited by laser pulses. The
tight binding approach is used for the band structure of CNT and the
effect of light is introduced by the vector potential. Contrary to the
common theoretical practice, no a priori assumptions are made concerning
electron wave packet, the latter is determined as a result of
illumination. In order to overcome the problem of various electron phases
in ZB, a method of two-photon echo is considered and described using the
density function formalism. The medium polarization of CNT is calculated
by computing exact solutions of the time-dependent electron Hamiltonian. The signal
of two-photon echo is extracted and it is shown that, using existing
parameters of CNT and laser pulses, one should be able to observe the
electron trembling motion. Effects of electron decoherence and relaxation
are discussed.
\end{abstract}

\maketitle

\section{Introduction}

The phenomenon of Zitterbewegung (ZB, trembling motion) was devised by Schrodinger in 1930 for free
electrons in a vacuum~\cite{Schroedinger1930}. Schrodinger showed that, in the Dirac equation, the
electron velocity operator does not commute with the Hamiltonian, so that the electron velocity is not
a constant of the motion also in the absence of external fields. Solving differential equation for
the time dependence of velocity Schrodinger demonstrated that, in addition to the classical motion,
the velocity contains a quickly oscillating Zitterbewegung component. For many years ZB was a subject
of theoretical fascination but also of controversy since many authors questioned its existence or
observability. It was later recognized that ZB is due to an interference between positive and negative energy
solutions of the Dirac equation~\cite{BjorkenBook}. In an important paper, Lock~\cite{Lock1979} showed
that, if an electron is represented by a wave packet, ZB oscillations decay in time as a consequence
of the Riemann-Lebesgue lemma.

Since 1970 the phenomenon of ZB was proposed for electrons in superconductors and semiconductors
in which, due to their energy spectra dominated by two bands separated by a gap, an interference of
positive and negative energy states can take place~\cite{Cremer1970,Cannata1990,Vonsovskii1990}.
In 1997 Zawadzki~\cite{ZawadzkiHMF} indicated
that, because of a close analogy between the behavior of electrons in narrow-gap semiconductors
and relativistic electrons in a vacuum, one can expect the trembling motion in narrow-gap semiconducting alloys
similar to that proposed by Schrodinger in a vacuum, but having much lower frequencies and
considerably higher amplitudes.

In 2005 papers by Zawadzki~\cite{Zawadzki2005} and Schliemann {\it et al.}~\cite{Schliemann2005} triggered
a real surge of theoretical works describing ZB in various systems,
see e.g.~\cite{Katsnelson2006,Zawadzki2006,Cserti2006,Rusin2007b,Clark2008,Demikhovskii2008,ZhangLiu2008,Romera2009,Rusin2009,Longhi2010b},
and the review~\cite{Zawadzki2011}. In particular, the present authors demonstrated that the ZB phenomenon in
a periodic potential can be regarded as a mode in which the electron keeps its total energy constant,
compensating the periodic potential changes by periodic changes of the kinetic energy~\cite{Zawadzki2010}.
A proof-of-principle simulation of the 1+1 Dirac equation with the resulting ZB of a wave packet was carried out by
Gerritsma {\it et al.}~\cite{Gerritsma2010} with the use of cold ions interacting with laser beams.

However, with all the theoretical progress made in the description of Zitterbewegung in solids, there
have been no experimental observations of the phenomenon. There are a few reasons for that. In recent
literature the electrons are generally represented in the form of Gaussian wave packets with large initial
quasi-momentum~$\hbar k_0$ transverse to the proposed direction of ZB, see e.g.~\cite{Schliemann2005,Rusin2007b}.
However, it is not clear how to prepare experimentally electrons in this form and, in particular, how to
transfer to an electron a large initial quasi-momentum. In a recent paper, the present authors indicated that
one should take an ``experimental'' approach, i.e. {\it not to assume} anything about the wave packet, but
to {\it determine it} from a known form of the experimental laser pulse that triggers
the whole ZB process~\cite{Rusin2013b}. Secondly, and this is probably the main reason for the
non-observation of ZB phenomenon in solids, the ``trembling electrons''
move in a crystal with different directions and phases, so that the oscillations may average to zero.
Thus, in order to observe the trembling directly, one would need to follow the motion of a single electron.

In order to overcome the above difficulties, we follow example of the Bloch oscillator (BO).
The Bloch oscillator is another phenomenon which had been proposed a long time ago and it took many years
to observe it. The phenomena of ZB and BO are basically similar, their nature is different, but they are both characterized by
electron oscillations with an inherent frequency: for BO it is determined by an external
electric field and for ZB by an energy gap between the conduction and valence bands. In both situations various
electrons oscillate with different phases. The Bloch oscillator was finally investigated and observed by means of nonlinear optics.
Von Plessen and Thomas~\cite{Plessen1992} used the third-order perturbation expansion in electric field to calculate a two-photon
echo (2PE) signal resulting from Bloch oscillations in superlattices. In this scheme the sample is illuminated by
two subsequent laser pulses with the wave vectors~${\bm k}_1$ and~${\bm k}_2$ at times~$t=0$ and~$t=\tau_D$, respectively.
Light is then emitted into the background-free direction~${\bm k}_3=2{\bm k}_2-{\bm k}_1$ due to the
nonlinear optical interaction in the sample~\cite{MukamelBook}. The above method, with some variations,
was successfully used to observe the Bloch oscillations in GaAs/GaAlAs superlattices, see e.g.~\cite{Feldmann1992,Lyssenko1997}.
The essential point that makes the techniques of nonlinear optics so powerful is the condition of
phase matching~${\bm k}_1+{\bm k}_2+{\bm k}_3=0$. When this condition is fulfilled,
the individual dipoles created by oscillating electrons are properly phased so that the field emitted by each dipole adds
coherently in the~${\bm k}_3$ direction, which strongly enhances the emitted signal~\cite{BoydBook}.

Following the path taken in above works we turn to the nonlinear optics and, more specifically,
to the two-photon echo technique~\cite{MukamelBook}. We show that such an experiment can be used to
detect the ZB phenomenon in solids. In our approach, we do not assume anything about the electron wave packet,
but determine it from excitation by a laser pulse. In calculating the medium polarization we do not use
perturbation expansions, but compute exact solutions from the time-dependent Hamiltonian.
We use the density matrix~(DM) formalism since we want to include effects of relaxation and decoherence.
Both these features affect the interference of positive and negative energy states which
underlines the trembling motion.

The paper is organized as follows. In Section~II we shortly summarize the band structure of CNT and
indicate how to introduce light to the formalism. Section~III contains the calculation of Zitterbewegung
and the resulting medium polarization. Section~IV concentrates on the two-photon echo and gives the results.
In Section~V we discuss the main features of our treatment. The paper is concluded by a summary.

\section{CNT Hamiltonian in laser field}

We first briefly summarize the tight-binding
approach to the energy bands in CNT and introduce the vector potential of laser field to the formalism.
Following Saito {\it et al.}~\cite{Saito1992,SaitoBook} we
consider two-dimensional hexagonal lattice in which carbon atoms are placed in two nonequivalent
points, called traditionally~$A$ and~$B$ points.
Each atom placed in the~$A$ point is surrounded by three atoms placed in the~$B$ points, whose relative positions
are:~${\bm R}_1=a/\sqrt{3}(1,0)$,~${\bm R}_2=a/\sqrt{3}(-1/2,\sqrt{3/2})$ and~${\bm R}_3=a/\sqrt{3}(-1/2,-\sqrt{3/2})$,
where~$a=2.46$~\AA\ is the length of carbon-carbon bond.
Within the usual tight-binding method one expands the electron Bloch function into a linear
combination of~$\phi_A$ and~$\phi_B$ atomic functions in~$A$ and~$B$ points.
The matrix elements of the periodic Hamiltonian~$\hat{\cal H}$ between atoms in two~$A$ or two~$B$ points
vanish, while the matrix element of~$\hat{\cal H}$ between atomic functions in~$A$ and~$B$ points is
\begin{equation} \label{ML_HAB1}
 {\cal H}_{AB}= \sum_{j=1}^3 t_j({\bm R}_j) e^{i{\bm k}\cdot {\bm R}_j},
\end{equation}
where~$t_j({\bm R}_j)$ are transfer integrals between the atom in~$A$ point and the atom in~$j$-th~$B$ point
\begin{equation} \label{ML_tRj}
 t_j({\bm R}_j) = \langle \phi_A({\bm r}) |{\cal H}| \phi_{Bj}({\bm r}-{\bm R}_j) \rangle.
\end{equation}
In the absence of fields there is for all~$j$:~$t_j({\bm R}_j)=t_{AB}$.
The tight-binding Hamiltonian of the graphene sheet is then
\begin{equation} \label{ML_hH}
 \hat{H}_M = \left(\begin{array}{cc} 0 & {\cal H}_{AB}^* \\ {\cal H}_{AB} & 0 \end{array} \right).
\end{equation}

A nanotube is obtained from a two-dimensional graphene sheet by rolling it into a cylinder.
As a result of folding, one joins lattice points connected
by the chiral vector~${\bm C}_h=n_1 {\bm a}_1 + n_2{\bm a}_2\equiv (n_1,n_2)$,
where~${\bm a}_1= a(3/2,\sqrt{3}/2)$ and~${\bm a}_2=a(3/2,-\sqrt{3}/2)$
are lattice vectors and~$n_1$,~$n_2$ are integers.
Here we consider a zigzag nanotube characterized by the chiral vector~${\bm C}_h=(N,0)$ for which, after the folding,
the wave vector~$k_y$ parallel to~${\bm C}_h$ becomes quantized:~$k_y=(2\pi/a)(m/N)$ with~$m=1 \ldots 2N$~\cite{Saito1992,SaitoBook}
and the~$k_x$ vector parallel to tube's axis remains continuous. Thus nanotube's band structure presents a set
of one-dimensional energy subbands. The tight-binding Hamiltonian for CNT can be obtained from Eq.~(\ref{ML_hH})
by replacing~${\cal H}_{AB}$ by
\begin{equation} \label{CNT_HAB2}
 {\cal H}_{AB}^T= e^{ika/\sqrt{3}} + 2e^{-ika/(2\sqrt{3})}\cos\left(\frac{m\pi}{N}\right),
\end{equation}
in which~$-\pi/a\sqrt{3} \leq k \leq \pi/a\sqrt{3}$, where we write~$k_x=k$.
For given~$k$, the energy subbands~$E_{k,m}$ in CNT obtained from Eqs.~(\ref{ML_hH}) and~(\ref{CNT_HAB2}) form~$4N$ subbands
symmetric with respect to~$E=0$, labeled by two quantum numbers:~$m=1 \ldots 2N$ and the energy sign~$\epsilon=\pm 1$.
For~$J=1 \ldots N-1$ each pair of energy subbands with~$m=N-J$ is degenerate with the pair having~$m=N+J$.
The subbands with~$m=N$ and~$m=2N$ are not degenerate. For other
properties of energy subbands in CNT, see~\cite{SaitoBook}.

We treat the laser light classically.
In the electric dipole approximation the field of a laser pulse is described
by the position-independent vector potential~${\bm A}={\bm A}(t)$
and the scalar potential~$\phi=0$.
To introduce the vector potential into the tight-binding formalism
we employ the method proposed by Graf and Vogl~\cite{Vogl1995}.
Following this approach we replace in Eq.~(\ref{ML_HAB1}) each~$t_j({\bm R}_j)$
by its potential-dependent counterpart~$T_j({\bm R}_j)$
\begin{equation} \label{ML_TRj}
 T_j({\bm R}_j) = t_j({\bm R}_j) \exp\left\{-\frac{ie}{2\hbar}{\bm R}_j\cdot[{\bm A}({\bm 0},t)+{\bm A}({\bm R}_j,t)]\right\}.
\end{equation}
The in-site energies are not modified:~${\cal H}_{AA}={\cal H}_{BB}=0$, because the scalar potential is zero.
Since~${\bm A}(t)$ does not depend on spatial coordinates, we have
\begin{equation} \label{ML_TRj2}
 T_j({\bm R}_j) = t_{AB} \exp\left\{-\frac{ie}{\hbar}{\bm R}_j \cdot {\bm A}(t) \right\},
\end{equation}
Then the time-dependent electron Hamiltonian is
\begin{equation} \label{CNT_hHt}
\hat{H}(t)= t_{AB} \left(\begin{array}{cc} 0 & h(t)^* \\ h(t) & 0 \end{array} \right),
\end{equation}
in which
\begin{equation} \label{CNT_h}
 h(t)= e^{iq(t)a/\sqrt{3}} + 2e^{-iq(t)a/(2\sqrt{3})}\cos\left(\frac{m\pi}{N}\right),
\end{equation}
and~${\bm q}(t)= {\bm k} -(e/\hbar){\bm A}(t)$ is the generalized quasi-momentum.
The final result of the approximations described above resembles the usual ``minimal coupling'' substitution for
the free-electron case in the presence of an electric field. This result is valid only
for vector potentials which do not depend on the spatial coordinates.

\section{Zitterbewegung in carbon nanotubes}

\begin{figure}
\includegraphics[width=8.0cm,height=8.0cm]{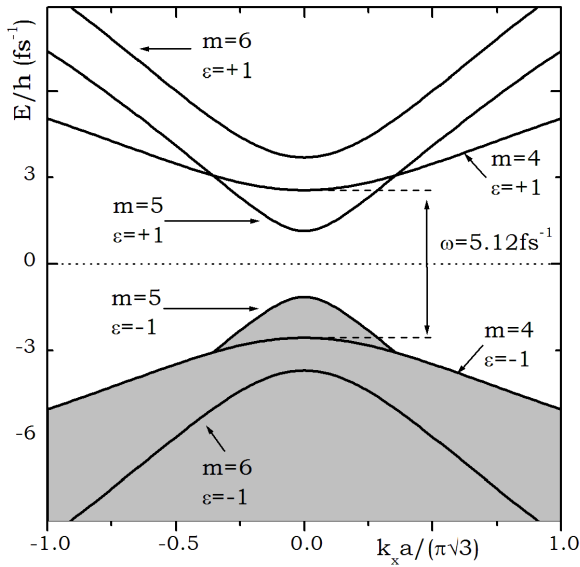}
\caption{Schematic plot of energy subbands close to~$E=0$ in zigzag CNT for parameters listed in
         Table~1, cf.~\cite{SaitoBook}.} \label{FigBands}
\end{figure}

\begin{table}
 \begin{tabular}{|c|l|l|}
   \hline

   $a$ & lattice constant & 2.46~\AA \\
   $k$ & wave vector &~$ka \in [-\pi/\sqrt{3},\pi/\sqrt{3}]$ \\
   $t_{AB}$ & matrix element for atoms & \\
            & placed in~$A$ and~$B$ points & 3.03 eV \\
   ${\bm C}_h$ & chiral vector~${\bm C}_h=(N,0)$ &~$(7,0)$ \\
   \hline
   $\omega_L$ & central laser frequency & 4.5~fs$^{-1}$ \\
   $\tau$ & common pulse length & 4.5~fs \\
   $\tau_D$ & delay between pulses &~$ 0\ldots 25$~fs \\
   $E_1, E_2$ & field intensities &~$4\times 10^9$~V/m \\
   $t_E$ & first pulse termination & 22~fs \\
   \hline
   $\omega_4$ &~$2E_{m=4}/\hbar$ & 5.12~fs$^{-1}$ \\
   $\omega_5$ &~$2E_{m=5}/\hbar$ & 2.28~fs$^{-1}$ \\
   $\omega_6$ &~$2E_{m=6}/\hbar$ & 7.39~fs$^{-1}$ \\
   \hline
   $T_1$ & relaxation time& 300~fs \\
   $T_2$ & decoherence time & 130~fs \\
   \hline
 \end{tabular}
\caption{Parameters for a zigzag carbon nanotube after~\cite{Saito1992,SaitoBook} (first box),
         laser pulses parameters after~\cite{Wirth2011} (second box),
         frequencies corresponding to the gap energies obtained from the tight-binding theory (third box),
         electron relaxation time (model parameter, see~\cite{Slavcheva2010})
         and empirical decoherence time after~\cite{Voisin2003}.}
\end{table}

We consider a single-walled zigzag semiconductor CNT with~$N=7$.
In Fig.~\ref{FigBands} we plot three lowest pairs of energy subbands of such nanotube.
In the following analysis we concentrate on pairs of subbands with~$m=4$
separated by the energy gap~$E_g/\hbar=5.12$~fs$^{-1}$, see Table~1.
We assume that the CNT is illuminated by two consecutive laser pulses producing electric fields
oscillating in the~$x$ direction. For simplicity we assume for the two pulses
the same pulse duration~$\tau$ and the central frequency~$\omega_L$. We assume that~$\omega_L$
is close to the interband frequency~$\omega_4=E_4/\hbar$, see Table~1.
For~$t \geq 0$ the electric field of the light is described by
\begin{equation} \label{Pulse_Et}
 {\bm E}(t) = \sum_{j=1,2}{\bm \epsilon}_jE_j\exp\left(-b\frac{t_j^2}{\tau^2}\right)\cos(\phi_j-\omega_Lt_j),
\end{equation}
while for~$t<0$ we take~$E(t)=0$. In the above expression~${\bm \epsilon}_j$
are beam polarizations,~$E_j$ are field intensities,~$t_1=t-t_0$, and~$t_2=t_1-t_D$ are relative times
of the first and second pulse, respectively,~$t_0=3\tau$ is the common shift of pulses
centers,~$t_D$ is the delay between pulses,
and~$b=2\ln2\simeq 1.386$. Phases~$\phi_j$ will be defined in Section IV.
The assumption of vanishing~${\bm E}(t)$ for~$t<0$ simplifies numerical
calculations of~${\bm A}(t)$, and for sufficiently large~$t_0$ it has no impact on final results.
For~$t\geq 0$ the vector potential corresponding to~${\bm E}(t)$ is
\begin{equation} \label{Pulse_At}
 {\bm A}(t) = - \int_0^{t}{\bm E}(t') dt' \hspace*{1em},
\end{equation}
while for~$t<0$ there is~${\bm A}(t)=0$. The potential~${\bm A}(t)$ is calculated numerically.
Since~${\bm A}(t)$ does not depend on spatial coordinates, the magnetic field of the wave
vanishes:~${\bm B}={\bm \nabla} {\bm \times} {\bm A}(t)=0$.

In the absence of fields, for a state characterized by~$k$ and~$m$, the Hamiltonian~$\hat{H}(0)$
has two eigenvectors~$w_1$ and~$w_2$ corresponding to the positive and negative
eigenenergies~$E_{k,m}=\pm t_{AB}|h(0)|$, respectively. There is
\begin{equation} \label{CNT_w1}
 w_1=\frac{1}{\sqrt{2}|h(0)|} \left(\begin{array}{c} |h(0)| \\ h(0)\end{array} \right),
\end{equation}
and
\begin{equation} \label{CNT_w2}
 w_2=\frac{1}{\sqrt{2}|h(0)|} \left(\begin{array}{c} -h^*(0) \\ |h(0)|\end{array} \right),
\end{equation}
see Eq.~(\ref{CNT_hHt}).
In the presence of dissipative processes the density matrix (DM)~$\hat{\rho}$ of the electron
evolves according to the Liouville equation~\cite{BoydBook}
\begin{equation} \label{Liou_H}
\frac{d}{dt}\hat{\rho} = \frac{-i}{\hbar}(\hat{H}\hat{\rho} -\hat{\rho}\hat{H}) - \hat{\cal D}(\hat{\rho}),
\end{equation}
where~$\hat{\cal D}(\hat{\rho})$ includes relaxation and decoherence processes taking place
during the electron motion. In the standard approach these processes are included phenomenologically
by assuming that the matrix elements of~$\hat{\cal D}(\hat{\rho})$
between states~$w_1$ and~$w_2$ are~\cite{BoydBook}
\begin{eqnarray} \label{Liou_Dump}
 \hat{\cal D}(\hat{\rho})_{ij} = \gamma_{ij} \langle w_i|\hat{\rho}-\hat{\rho}^{eq}|w_j\rangle \hspace*{1em} (i,j=1,2),
\end{eqnarray}
where~$\hat{\rho}^{eq}$ is the DM at equilibrium. We assume that in the absence of light
all subbands with negative energies are occupied and those with positive energies are empty.
Therefore~$\rho^{eq}_{22}=1$ and all other~$\rho^{eq}_{ij}=0$.
The constants~$\gamma_{ij}$ are:~$\gamma_{11}=\gamma_{22}=1/T_1$, where~$T_1$ is the relaxation time
of electron population excited to the conduction subbands. Experimental values of~$T_1$
are not well known, see~\cite{Slavcheva2010}. Further,~$\gamma_{12}=\gamma_{21}=1/T_2$, where~$T_2$
characterizes the decay of coherence between positive and negative energy states.
The decoherence time~$T_2$ in CNT was determined experimentally in Ref.~\cite{Voisin2003}, see Table~1.
We disregard long-time relaxation processes with~$T > 2000$~fs, as observed in CNTs~\cite{Kono2003}.
Since~$\rho_{12}=\rho_{21}^*$, the time evolution of four DM components,
as given in Eqs.~(\ref{Liou_H}) and~(\ref{Liou_Dump}), is~\cite{BoydBook}
\begin{widetext}
\begin{eqnarray}
 \label{Liou_rho11}
\frac{d}{dt}\rho_{11} &=& -\frac{1}{T_1}\rho_{11} - 2{\rm Im}(H_{21}) {\rm Re}(\rho_{21}) + 2{\rm Im}(\rho_{21}){\rm Re}(H_{21}), \\
 \label{Liou_rho22}
\frac{d}{dt}\rho_{22} &=& +\frac{1}{T_1}\rho_{11} + 2{\rm Im}(H_{21}) {\rm Re}(\rho_{21}) - 2{\rm Im}(\rho_{21}){\rm Re}(H_{21}), \\
 \label{Liou_Rerho21}
\frac{d}{dt}{\rm Re}(\rho_{21}) &=& -\frac{1}{T_2} {\rm Re}(\rho_{21}) + {\rm Im}(\rho_{21})(H_{22}-H_{11}) + {\rm Im}(H_{21})(\rho_{11}-\rho_{22}), \\
 \label{Liou_Imrho21}
\frac{d}{dt}{\rm Im}(\rho_{21}) &=& -\frac{1}{T_2} {\rm Im}(\rho_{21}) + {\rm Re}(\rho_{21})(H_{11}-H_{22}) + {\rm Re}(H_{21})(\rho_{22}-\rho_{11}).
\end{eqnarray} \end{widetext}
In the above equations,~$H_{ij}$ denote the matrix elements of the Hamiltonian~$\hat{H}(t)$ between
the states~$w_i$ and~$w_j$, see Eqs.~(\ref{CNT_w1}) --~(\ref{CNT_w2}). One should note that~$d\rho_{11}/dt+d\rho_{22}/dt=0$,
which ensures the conservation of electron density probability.
In the absence of fields,~$w_1$ and~$w_2$ are the eigenstates of~$\hat{H}(0)$ and then~$H_{12}=H_{21}=0$.
In this case one obtains three separate first-order equations for~$\rho_{11}$,~$\rho_{22}$ and~$\rho_{21}$, which can be solved
analytically. For nonzero electric fields, equations~(\ref{Liou_rho11}) --~(\ref{Liou_Imrho21})
have to be solved numerically. To specify the initial conditions we again assume that at~$t=0$,
when the electric field is not turned on yet,
all valence subbands are occupied and all conduction subbands are empty.
This gives:~$\rho_{22}(0)=1$ and all other~$\rho_{ij}(0)=0$ for every subband index~$m$ and wave vector~$k$.

The velocity operator in the~$x$ direction is
\begin{equation}\label{Liou_hv}
\hat{v} = \frac{\partial \hat{H}(t)}{\partial \hbar k},
\end{equation}
with~$\hat{H}(t)$ given in Eq.~(\ref{CNT_hHt}). The operator~$\hat{v}$
is represented in the of upper and lower components of the tight-binding function, see Eq.~(\ref{ML_hH}).
We introduce an operator~$\hat{V}$, which is the counterpart of~$\hat{v}$
in the basis of~$w_1$ and~$w_2$ states. The matrix elements of~$\hat{V}$ are
\begin{equation} \label{Liou_hV}
 \langle i|\hat{V}|j\rangle = \langle w_i\left|\frac{\partial \hat{H}(t)}{\partial \hbar k} \right| w_j\rangle.
\end{equation}
Then, for a state characterized by~$k$ and~$m$, the average electron velocity is
\begin{equation} \label{Liou_vkxmy}
 \langle v^{k,m}(t) \rangle= {\rm Tr} \{\hat{\rho}(t) \hat{V}\}.
\end{equation}
The total electron velocity integrated over~$k$ and summed over the subbands is
\begin{equation} \label{Liou_v}
 \langle v(t)\rangle=\frac{1}{2\pi}\sum_{m=1}^{2N} \int_{-k_{max}}^{k_{max}} \langle v^{k,m}(t) \rangle dk,
\end{equation}
where the Brillouin zone boundary is~$k_{max}=\pi/(a\sqrt{3})$~\cite{Saito1992,SaitoBook}.
The average packet position is:~$\langle x(t)\rangle = \int_0^t \langle v(t')\rangle dt'$
with~$\langle x(0)\rangle=0$. Then, the medium polarization~$P(t)$ induced by the laser light is
\begin{equation} \label{Liou_P}
P(t) = -|e|\langle x(t)\rangle = -|e|\int_{0}^{t} \langle v(t') \rangle dt',
\end{equation}
in which we assumed that at~$t=0$ the polarization vanishes.
We also calculate a probability~${\cal P}^-(t)$ of finding the electron in the states
with negative energies
\begin{equation} \label{Liou_pm_kxmx}
 {\cal P}^-(t)= \frac{1}{2\pi}\sum_{m=1}^{2N} \int_{-k_{max}}^{k_{max}} \rho_{22}^{k,m} dk.
\end{equation}
For~$t=0$ there is~${\cal P}^+_{k,m}(t)=0$ since we assumed that electrons occupy the valence states only.
The necessary condition for an appearance of ZB is that both probabilities~${\cal P}^{\pm}_{k,m}(t)$ do not vanish,
which is achieved by excitation of electrons with the use of laser pulses.

\begin{figure}
\includegraphics[width=8.0cm,height=8.0cm]{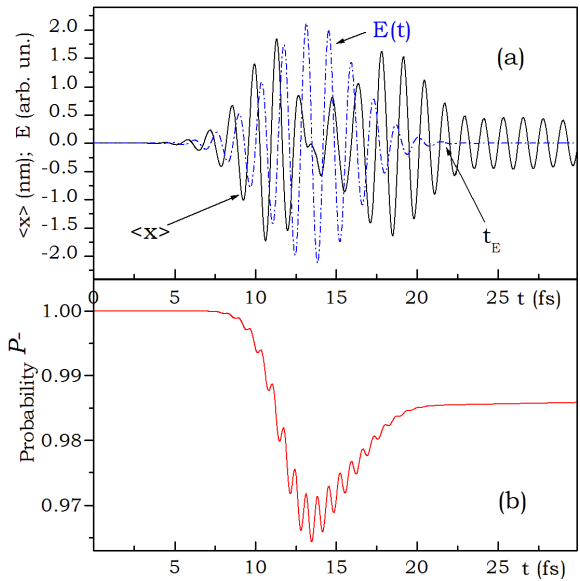}
\caption{(a) Average packet position versus time calculated for CNT and single laser pulse taking parameters
         listed in Table~1 with~$E_2=0$. Solid line: packet position,
         dashed line: electric field of the laser pulse (in arbitrary units).
         Arrow indicates time~$t_E$ at which the electric field of the pulse vanishes.
         (b) Normalized probability~${\cal P}^-$ of finding the electron in states with negative energy versus time.
        }\label{Fig2}
\end{figure}

The position~$\langle x(t)\rangle$~is proportional to the medium polarization induced
by the laser light, see Eq.~(\ref{Liou_P}). In Fig.~2(a) we plot oscillations of 
the packet position~$\langle x(t)\rangle$,
induced by a single laser shot (i.e., for~$E_2=0$) versus time, as calculated for the parameters listed in Table~1.
Pulses characterized by such parameters were obtained experimentally, see~\cite{Wirth2011}.
In the same figure we show the electric field of the laser pulse vanishing around~$t_E\simeq 22$~fs.
It is seen that the electron oscillates also in the absence of electric field,
so that the system remains polarized a long time after the pulse termination.
The polarization oscillates with one frequency, whose period~$T_Z=1.25$~fs
corresponds to the interband frequency~$2\pi/T_Z \simeq \omega_4=5.12$~fs$^{-1}$.
Therefore, for~$t\ge t_E$ the electron motion can be identified as the Zitterbewegung.
In Fig.~2(b) we show the probability of finding the electron in states of negative energies.
At the outset, the electrons are in the valence subbands, then the laser pulse excites small
fraction of electrons to the conduction subbands and after the pulse termination
the created wave packet consists of states having positive and negative energies.
The obtained wave packet is determined only by the material constants and pulse parameters. Thus,
there are no {\it a priori} assumptions concerning packet characteristics, in
contrast to previous works, see e.g.~\cite{Schliemann2005,Rusin2007b,Zawadzki2011}.
We note in passing that results similar to those shown in Fig.~2 were obtained
in Ref.~\cite{Rusin2013b} for a metallic CNT with~$N=9$.

In order to describe the packet motion after the pulse
termination we solve Eqs.~(\ref{Liou_rho11}) --~(\ref{Liou_Imrho21})
taking~${\bm E}(t)=0$ for~$t\geq t_E$, and
obtain:~$\rho_{11}(t)=\rho_{11}(t_E)e^{-(t-t_E)/T_1}$,~$\rho_{22}(t)= 1-\rho_{11}(t)$, and
\begin{equation}
\rho_{21} = \rho_{21}(t_E) \exp\left\{\left(-\frac{1}{T_2} + i\omega_4\right)\left(t-t_E\right)\right\},
\end{equation}
in which~$\rho_{11}(t_E)$ and~$\rho_{21}(t_E)$ are the DM elements at~$t=t_E$,
while~$\omega_4$ is given in Table~1. The results are valid for~$t\geq t_E$.
As pointed out in Ref.~\cite{Rusin2014a}, the ZB motion is related to the nondiagonal elements
of the density matrix, so that in the field-free case the velocity (or polarization) oscillates
with the interband frequency~$\omega_4$. This is clearly visible in Fig.~2(a).
The oscillations decay exponentially with the characteristic time~$T_2=130$~fs.
The electron relaxes from conduction to the valence subbands with the characteristic
time~$T_1=300$~fs, not visible in Fig.~2(b).

\section{Two-photon echo}

The above results suggest that, by measuring the time-dependent medium polarization, one can
directly observe electron ZB oscillations in CNT.
However, in real systems one deals with {\it many} electrons, which are excited with random phases
depending on their initial phases as well as on collisions with other electrons or lattice defects.
Thus, the polarization created by the laser shot will be destroyed and no net polarization will be measured.
As mentioned in the Introduction, the way to overcome this difficulty is to use nonlinear
laser spectroscopy~\cite{MukamelBook}, in which
two or more laser pulses allow one to observe nonlinear polarization components and to select signals with
proper phase-matching conditions~\cite{BoydBook}. Common methods to measure electron coherence are the two-photon echo (2PE)
and the degenerate four-wave mixing spectroscopies (DFWM). In the following we calculate the 2PE signal corresponding
to the ZB oscillations induced by two laser pulses.

The configuration of the two-photon echo experiment is described in Ref.~\cite{Yajima1979}.
We consider two incident laser beams characterized by two non-parallel wave vectors~${\bm k}_1$ and~${\bm k}_2$.
The first laser pulse creates the medium polarization [as shown in Fig.~2(a)],
which propagates in time after pulse disappearance. The second pulse probes the electron state and
leads to a coherent emission of the signal.
In the homodyne detection scheme the two-photon echo signal is measured in
the background-free direction~$2{\bm k}_2-{\bm k}_1$. The signal intensity~$S_{\tau_D}$
depends on the delay between two pulses, see Eq.~(\ref{Pulse_Et}),
and on the polarization component~$\widetilde{P}_{2{\bm k}_2-{\bm k}_1}(t)$ in the~$2{\bm k}_2-{\bm k}_1$
direction. Then there is~\cite{MukamelBook}
\begin{equation} \label{2PE_S}
 S_{\tau_D} \propto \int_{-\infty}^{\infty}\left|\widetilde{P}_{2{\bm k}_2-{\bm k}_1}(t)\right|^2 dt.
\end{equation}
To extract the polarization~$\widetilde{P}_{2{\bm k}_2-{\bm k}_1}(t)$ from the total polarization~$P(t)$,
as given in Eq.~(\ref{Liou_P}), we apply a non-perturbative method proposed
by Seidner and coworkers~\cite{Seidner1995,Seidner1997}.
In this method one introduces two auxiliary phases~$\phi_1$ and~$\phi_2$ to the electric
fields of the first and the second laser pulse, respectively,
as given in Eq.~(\ref{Pulse_Et}). Then, by solving Eqs.~(\ref{Liou_rho11}) --~(\ref{Liou_Imrho21}) one obtains
first a phase-dependent density matrix~$\hat{\rho}(\phi_1,\phi_2)$,
and then the polarization~$P(t,\phi_1,\phi_2)$, see Eqs.~(\ref{Liou_vkxmy}) --~(\ref{Liou_P}).
As it was shown in Refs.~\cite{Seidner1995,Seidner1997} within the rotating wave approximation (RWA),
there is
\begin{equation} \label{2PE_NL}
 \widetilde{P}_{2{\bm k}_2-{\bm k}_1}(t) =
 \frac{1}{2} \left[ \widetilde{P}\left(0\right) -i\widetilde{P}\left(\frac{\pi}{2}\right)
 -\widetilde{P}\left(\pi\right) + i\widetilde{P}\left(\frac{3\pi}{2}\right) \right],
\end{equation}
in which the notation~$\widetilde{P}\left(\phi \right) \equiv \widetilde{P}\left(t,\phi,0\right)$ is used.
The tilde-polarization is~\cite{Seidner1997,Manal2006}
\begin{equation} \label{2PE_Ptilde}
 \widetilde{P}(\phi_1)= P(t,\phi_1,0) -P_{E_1}(t,\phi_1) - P_{E_2}(t,\phi_1),
\end{equation}
where~$P_{E_1}(t,\phi_1)$,~$P_{E_2}(t,\phi_1)$ describe polarizations calculated taking
into account only the first or only the second laser pulse, respectively,
with the same phase~$\phi_1$ of electric fields.
The above approximation is valid up to all orders of electric field~\cite{Seidner1995},
but it is limited to the range of validity of RWA, i.e., to the laser frequencies
close to the interband frequency~$\omega_Z$. In our case this condition is fulfilled, see Table~1.

Dependence of the intensity~$S_{\tau_D}$ in Eq.~(\ref{2PE_S}) on the delay~$\tau_D$ is
calculated numerically in a few steps.
First, we select a value of~$0 \leq \tau_D \leq 25$~fs and for this~$\tau_D$
we choose one of four values of~$\phi_1$, see Eq.~(\ref{2PE_NL}).
Then we select~$2M+1$ mesh points of~$k_{max} \leq k \leq k_{max}$, for each of~$1 \leq m \leq 2N$ pairs
of subbands. We use~$M=400$,~$N=7$, and~$k_{max}=\pi/(a\sqrt{3})$.
Next, for all~$k$ and~$m$, we solve Eqs.~(\ref{Liou_rho11}) --~(\ref{Liou_Imrho21})
using the fourth order Runge-Kutta method. For the CNT parameters listed in Table~1
the period of interband oscillations varies from~$0.3$~fs to~$2.8$~fs.
In numerical calculations we use time-step~$t_s\simeq 0.01$~fs, much
smaller than~$0.3$~fs. We terminate the calculation at~$t_{max}=100$~fs. Further,
we calculate the total polarization~$P(t,\phi_1,0)$ in Eq.~(\ref{Liou_P})
and repeat the above procedure for
three remaining values of~$\phi_1$, see Eq.~(\ref{2PE_NL}).
Having calculated~$P(t,\phi_1,0)$ for four values of~$\phi_1$, we
calculate~$\widetilde{P}_{2{\bm k}_2-{\bm k}_1}(t)$ in Eq.~(\ref{2PE_NL})
and the intensity~$S_{\tau_D}$ in Eq.~(\ref{2PE_S}). The integration in Eq.~(\ref{2PE_S}) is
truncated at~$t_{max}=100$~fs. In this step we need~$P_{E_1}(t,\phi_1)$ and~$P_{E_2}(t,\phi_1)$,
as given in Eq.~(\ref{2PE_Ptilde}), which have to be precalculated earlier using the same approach,
but taking~$E_2=0$ or~$E_1=0$, respectively.
Next, we select another value of~$\tau_D$ with the time-step~$0.1$~fs and repeat the whole procedure.

\begin{figure}
\includegraphics[width=8.0cm,height=8.0cm]{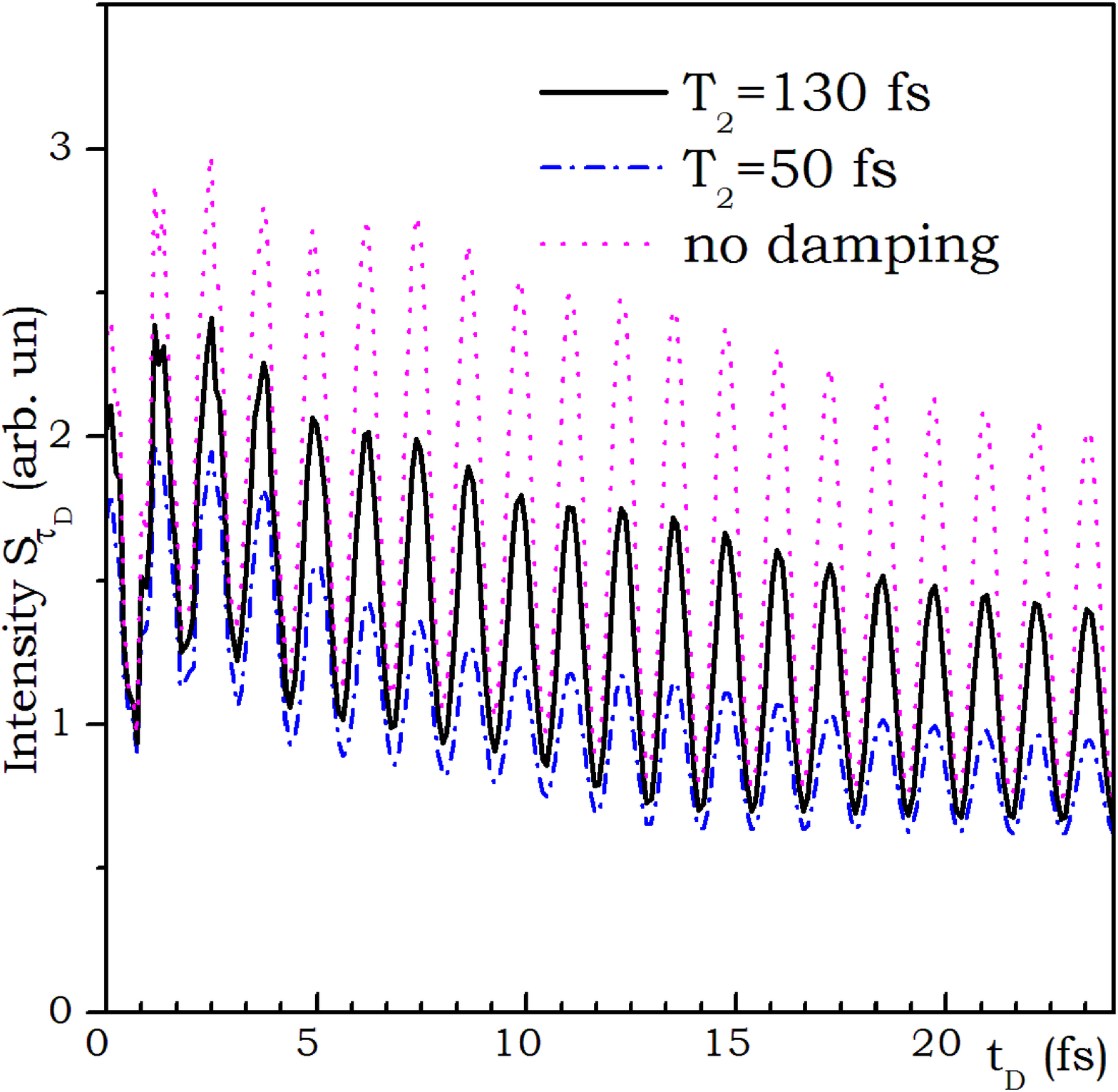}
\caption{Time-integrated two-photon echo signal~$S_{\tau_D}$ as given in Eq.~(\ref{2PE_S})
         calculated versus delay between the two pulses using non-perburbative method.
         Upper dotted line: damping excluded; solid line: damping times listed in Table~1,~$T_2=130$~fs
         obtained experimentally in~\cite{Voisin2003}, lowest dashed line: short artificial
         damping times~$T_1=T_2=50$~fs. The ZB oscillations correspond to~$\tau_D > 13.5$~fs, see text.
        }\label{Fig3}
\end{figure}

\begin{figure}
\includegraphics[width=8.0cm,height=8.0cm]{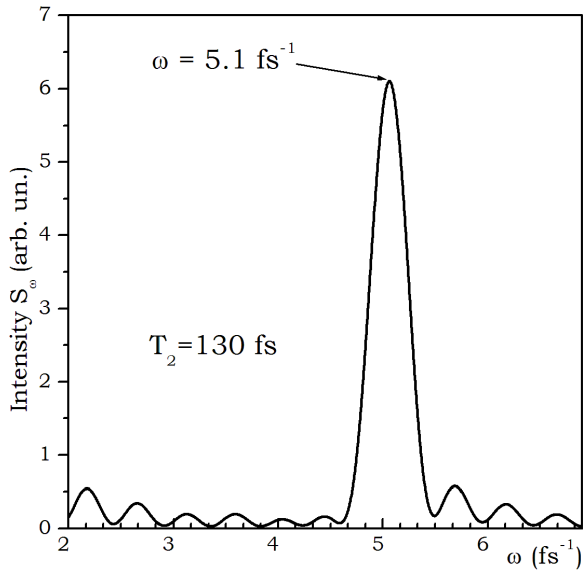}
\caption{Fourier transform of signal plotted in Fig.~3 by the solid line.
         The pronounced maximum at~$\omega=5.1$~fs$^{-1}$ corresponds to Zitterbewegung.}\label{Fig4}
\end{figure}

The results of our calculations are presented in Fig.~3, where we plot the signal~$S_{\tau_D}$ as a
function of~$\tau_D$ for three sets of dephasing times. The dotted line corresponds to the no-damping case,
the solid line corresponds to the relaxation time~$T_1$ and experimental decoherence time~$T_2$ listed in Table~1,
while the dashed line corresponds to artificially short times~$T_1=T_2=50$~fs.
The common feature of the three results shown in Fig.~3 is their oscillating character with the same period~$T_Z\simeq 1.25$~fs,
corresponding to the frequency~$\omega_Z=2\pi/T_Z = 5.03$~fs$^{-1}$. This value is close to the
interband frequency~$\omega_4$, see Fig.~1. The amplitudes of the signals depend on~$T_1$ and~$T_2$, but in all
cases the oscillations are clearly visible. This indicates that the Zitterbewegung is a robust phenomenon,
not sensitive to the details of model parameters. Figure~3 shows the main result of our work.

In Fig.~4 we show the power spectrum of the signal~$S_{\omega}$,
which is calculated by taking the Fourier transform of~$S_{\tau_D}$ corresponding to the solid line in Fig.~3.
The maximum of~$S_{\omega}$ is close to the interband frequency~$\omega_4=5.12$~fs$^{-1}$. This confirms that the
signal~$S_{\tau_D}$ oscillates with the ZB frequency~$\omega_Z$,
so that the medium polarization exhibits oscillations with the interband frequency.
As discussed in Ref.~\cite{Rusin2013b}, the ``true'' ZB oscillations
are represented by the field-free electron motion, which occurs after termination of the first laser pulse.
For this reason the oscillations of the intensity~$S_{\tau_D}$ truly monitor the electron ZB oscillations only
for~$\tau_D \gtrsim 3\tau = 13.5$~fs.

To verify accuracy of our results shown in Fig.~3 we calculate the intensities~$S_{\tau_D}$
assuming a longer cut-off time~$t_{max}=200$~fs. In this case the signals~$S_{\tau_D}$
are similar to those presented in Fig.~3. For~$T_1=T_2=50$~fs (dashed line) the change of the cut-off
time alters the results by less than~$0.5\%$, for~$T_1$,~$T_2$ listed in Table~1 the results
are changed by~$\simeq 3.0\%$, while for the damping-free case (dotted line)
they differ by around~$16\%$. This analysis confirms validity of the results plotted
in Fig.~3 for finite damping times. On the other hand, the damping-free
results strongly depend on the cut-off time.

The results shown in Fig.~3 indicate a possibility of the experimental observation of ZB
motion with the use of the two-photon echo experiment. Each element of such
an experiment is available within the current techniques.
Carbon nanotubes characterized by parameters
listed in Table~1 were created many years ago~\cite{SaitoBook}. Laser pulses considered in
our description were obtained experimentally in Ref.~\cite{Wirth2011}.
A desired separation~$\tau_D$ between two pulses, being on the order of~$0.1$~fs, was reported
in Ref.~\cite{Goulielmakis2004} for pulses with similar frequencies and field intensities to those
listed in Table~1. Experimental observations of photon echo in CNT were carried out, 
for example, in Ref.~\cite{Goulielmakis2004}.
All in all, it seems possible to observe the ZB oscillations
using the method proposed in the present work.
We hope that the present approach will motivate experiments
detecting the phenomenon of electron trembling motion in solids.

\section{Discussion}

It is important to find proper material and pulse parameters allowing for an observation of ZB
oscillations. First, the spectrum of both laser pulses should contain frequencies corresponding to
interband frequencies between at least one pair of energy subbands.
Next, the central laser frequency~$\omega_L$
should be close, but not equal to the selected interband frequency. Further,
too wide spectra of the pulses result in an excitation of electron motion
with many interband frequencies which makes it difficult to observe
and interpret the 2PE signal~\cite{Rusin2013b}. The above requirements are fulfilled
for light pulses having the width of a few cycles. Finally, the ZB oscillations of electron position
must exist after pulse termination, see Fig.~2(a). The material and pulse parameters listed in Table~1
fulfill all above conditions. It is to be noted that for pulses obtained
using a Ti:Sapphire laser tuned to~$\omega_5=2.28$~fs$^{-1}$, the last condition seems to be not met.

In this work we apply a non-perturbative calculation of the two-photon echo signal with the use of
the vector-potential gauge. We obtain the total polarization~$P(t)$
via the average velocity, see Eqs.~(\ref{Liou_vkxmy}) --~(\ref{Liou_P}).
In standard procedures employed in nonlinear optics,
the electric-field gauge is used which automatically ensures the gauge invariance of
the results~\cite{MukamelBook}. In or approach we apply the vector-potential gauge because
then the wave-vector~$k$ in the Hamiltonian~(\ref{CNT_hHt}) is a good quantum number and
equations~(\ref{Liou_rho11}) --~(\ref{Liou_Imrho21}) are ordinary differential equations. In the
electric-field gauge the corresponding equations would be partial differential equations, more
difficult to solve. Still, the use of the vector potential
gives correct results for the following reasons.
First, the Liouville equation (\ref{Liou_H}) without the damping term is gauge-invariant.
Second, as pointed out in Refs.~\cite{Tokman2009,Lamb1987},
the phenomenological damping term~$\hat{\cal D}(\hat{\rho})$ in Eq.~(\ref{Liou_H}) is gauge-invariant
in the dipole approximation, if the characteristic length of the problem is much smaller that the
light wavelength. In our system this condition is satisfied because the laser wavelength is much larger
than the CNT radius, see~\cite{SaitoBook}. The perturbation expansion within the vector-potential gauge
may be incorrect if one truncates the expansion at a finite order terms~\cite{Kobe1978}.
In our case the numerical calculation of the polarization~$P(t,\phi_1,\phi_2)$ ensures that
the vector potential is taken into account in {\it all} orders of the perturbation series,
so that the gauge invariance is preserved. The nonlinear polarization~$\widetilde{P}_{2{\bm k}_2-{\bm k}_1}(t)$
is extracted from~$P(t,\phi_1,\phi_2)$ in the non-perturbative way using the RWA~\cite{Seidner1995},
so that it is also gauge-invariant. This is also true for~$S_{\tau_D}$.
Therefore, the use of vector-potential gauge requires an application of the non-perturbative way to
calculate the polarization~$\widetilde{P}_{2{\bm k}_2-{\bm k}_1}(t)$.
On the other hand, the calculation of the third-order polarization~$\widetilde{P}_{2{\bm k}_2-{\bm k}_1}^{(3)}(t)$,
carried out in quantum chemistry~\cite{MukamelBook}, has to be performed in the electric-field gauge,
since the perturbation series is truncated.

In our description we use the density matrix formalism.
The DM approach allows one to take into account dephasing processes.
Since we consider two laser shots separated by the delay~$\tau_D$, the decoherence time~$T_2=130$~fs may
not be neglected, especially for large~$\tau_D$. The time integration in Eq.~(\ref{2PE_S}) extends to
large temporal values and the presence of damping ensures its convergence.
In the absence of damping the polarization~$|\widetilde{P}_{2{\bm k}_2-{\bm k}_1}(t)|^2$ in Eq.~(\ref{2PE_S})
decays as~$t^{-\alpha}$ with~$\alpha \gtrsim 1$, which may lead to problems
with the convergence of integrals for large times.
In the presence of dephasing processes the integrand vanishes exponentially~\cite{Rusin2014a}.
Finally, in the absence of relaxation processes, the electron
remains forever in the state with small admixture of positive energy states
while, in the presence of damping, it relaxes to the ground state with the
characteristic time~$T_1 \simeq 300$~fs, which makes the description more realistic.

The 2PE signal is obtained from the total polarization~$P(t)$ induced by the laser shots which,
in this work, is calculated within the DM formalism. In principle,
however,~$P(t)$ can be obtained using simpler methods, e.g. by solving the Schrodinger equation with
the time-dependent Hamiltonian~\cite{Rusin2013a,Rusin2013b}, in which damping effects are neglected.
As a matter of example, the results indicated by the dotted line in Fig.~3, correspond to the damping-free case,
as obtained using the wave-function approach with the time-dependent Hamiltonian (\ref{CNT_hHt}).
It is seen that the damping-free results have the same period and phase as those
calculated for finite damping times. On the other hand, the total polarization can be obtained appying
more sophisticated methods like the CNT Bloch equations~\cite{Hirtschulz2008,Malic2008}.
The CNT Bloch equations allow one to obtain~$P(t)$ taking into account the existence of excitons
and many-body effects like a renormalization of subbands energies. But
the presence of excitons and many-body effects weakly alters the energy gap
of CNT, so that the resulting polarization would oscillate with a frequency
close to that shown in Figs.~3 and~4.

As mentioned in the Introduction, von Plessen and Thomas~\cite{Plessen1992}
used similar approach to calculate the two-photon signal
for coherently exited Bloch oscillations in superlattices. In their work,
the 2PE was calculated using perturbation expansion of
the time-dependent electron wave function. The signal was computed within the third order of
the incident field and it exhibited oscillations with the period equal to that of the Bloch oscillations.
In our approach we propose a similar method but we
apply different formalism for the reasons given above.

The Degenerate Four Wave Mixing (DFWM) technique was successfully applied for experimental observation
of Bloch oscillations~\cite{Feldmann1992}. This method would be also applicable for an observation of ZB oscillations.
In DFWM one uses three laser pulses having two different frequencies~\cite{MukamelBook}.
The theory of DFWM in a non-perturbative approach was described in Ref.~\cite{Meyer2000}.
The spectroscopic signal may be calculated from twelve combinations of laser beam phases
by formulas analogous to those in Eq.~(\ref{2PE_NL}),
which requires a considerably larger numerical effort than that for 2PE, as presented here.
The method of DFWM seems to be less practical for our system since it requires three laser pulses instead of two.
But if this complication is of minor importance, DFWM might be also applied to observations of ZB.

In the present work we concentrate on zigzag nanotubes since for such CNTs the subbands minima and maxima
occur at~$k=0$. This feature simplifies the model and the calculations but it is not decisive
for the existence of ZB. In our considerations we chose
the semiconductor zigzag CNT with~$N=7$. However,
the results similar to those shown in Fig.~3 can be obtained for other~$N$ values as well as for
other CNT types: arm-chair and chiral ones~\cite{SaitoBook}.
The ZB oscillations should be observable by 2PE experiments also in super-lattices.
The crucial point is to adjust the laser central frequency~$\omega_L$ to the energy gap of the material.

The results shown in Fig.~3 require a fairly large computational effort, which is the main disadvantage of the
non-perturbative calculation of 2PE signals. However, there are several practical advantages of this approach.
First, once the numerical procedures are tested and implemented, they can be equally well applied to low and high
electric fields, as well as to overlapping or non-overlapping pulses of arbitrary length.
Another advantage is that similar calculations can be repeated for different configurations of phases, 
mesh points etc. But the most important point is that the non-perturbative procedure ensures 
gauge-invariant results within the vector-potential gauge.

\section{Summary}

We considered the electron Zitterbewegung in zigzag carbon nanotubes
excited by laser light. The band structure of CNT was described within the
tight binding approach and the light was introduced to the electron
Hamiltonian by the vector potential in the electric dipole approximation.
In contrast to common treatments, we did not make any assumptions
concerning the electron wave packet, but determined it as a result of
illumination by light. Trying to solve the problem created by the fact
that the trembling electrons move in various directions and their ZB
oscillations have various phases, we considered a two-photon echo method
successfully used to detect the Bloch Oscillator phenomenon. The medium
polarization of CNT was calculated in the density function formalism by
finding numerical solutions of the time-dependent Hamiltonian for
electrons in CNT. The two-photon echo signal was then extracted from the total
medium polarization and it was shown that it is possible to perform the
two-photon echo experiment with the existing CNT parameters and laser
pulse characteristics. Such an experiment should unambiguously detect the
phenomenon of Zitterbewegung in solids.

\end{document}